
\font\tsnorm=cmr10

\magnification=1200
\def\aa{{\alpha}}

\def\dd{{\delta}}
\def\ee{{\epsilon}}

\def\ll{{\lambda}}

\def\gg{{\gamma}}
\def\tt{{\tau}}

\def\ss{{\sigma}}
\def\SS{{\Sigma}}

\def\vl{\nabla\hskip -1pt \cdot \hskip -1 pt  {\rm L}}
\def\kl{{\rm L}}
\def\ref #1{\medskip\item{#1}}
\def\beginsection #1 #2
{\bigskip\bigskip \noindent{\bf #1 \  \  #2} \nobreak\noindent}
\hsize=150truemm
\vsize=8.2truein
\hoffset=6truemm
\voffset=0.25truein
\topskip=25pt
\null
\vskip 5truemm
\vskip 15truemm
\centerline{\bf Asymptotically Hyperbolic}
\centerline{\bf Non Constant Mean Curvature Solutions }
\centerline{\bf of the Einstein Constraint Equations}
\vskip 20truemm

\centerline{\tsnorm James Isenberg}
\centerline{\tsnorm {   jim@newton.uoregon.edu}}
\centerline{\tsnorm Department of Mathematics}
\centerline{\tsnorm and}
\centerline{\tsnorm Institute for Theoretical Sciences}
\centerline{\tsnorm University of Oregon}
\centerline{\tsnorm Eugene, OR 97403}
\centerline{\tsnorm USA}
\bigskip\bigskip
\centerline{\tsnorm Jiseong Park}
\centerline{\tsnorm park@aei-potsdam.mpg.de}
\centerline{\tsnorm Max-Planck-Institut f\"ur Gravitationsphysik}
\centerline{\tsnorm Albert-Einstein-Institut}
\centerline{\tsnorm D-14473, Potsdam}
\centerline{\tsnorm Germany}

\vskip 0.9in
\centerline{\bf Abstract}
\bigskip

We describe how the iterative technique used by Isenberg and Moncrief
to verify the existence of large sets of  non constant mean curvature
solutions of the Einstein constraints on closed manifolds can be adapted
to verify the existence of large sets of asymptotically hyperbolic
non constant mean curvature solutions of the Einstein constraints.

\vfil
\eject

     
\beginsection \S1 Introduction
\bigskip

For many years, the Einstein constraint equations have been
studied primarily on either
closed manifolds or on open manifolds with asymptotically Euclidean boundary
conditions. Such a concentration makes sense if one focuses on the 
Cauchy problem for cosmological spacetimes or on the Cauchy problem for
asymptotically flat spacetimes in a neighborhood of a Cauchy surface
which goes to spacelike infinity.

Recent work of Friedrich [11] has shown that one can very usefully study
asymptotically flat spacetimes using a Cauchy problem based on  spacelike
hypersurfaces which approach null infinity rather than spacelike infinity.
The prototype for such hypersurfaces is the ``constant mass" hyperboloid
hypersurface in Minkowski spacetime which has constant negative intrinsic 
curvature. More generally, these spacelike hypersurfaces do not have 
constant negative curvature, but they necessarily approach 
(at least locally) a constant
negative curvature hypersurface asymptotically. Hence, in studying the
constraint equations on hypersurfaces of this sort, one imposes 
asymptotically hyperbolic boundary conditions (rather than the more familiar 
asymptotically Euclidean boundary conditions). We shall detail these below.

Regardless of the topology or the boundary conditions on the hypersurface,
the constraint equations are much simpler to study for initial data with 
constant mean curvature (CMC) than for non CMC initial data. This is 
because, in the CMC case, three of the four constraint equations are
essentially trivial, and so one need only work with one nonlinear partial
differential equation; while in the non CMC case, there are four coupled
PDEs which must be handled. As a consequence of this difference, while 
CMC solutions of the constraint equations are essentially fully understood
(on closed manifolds
[9] [12], for asymptotically Euclidean data [6] [7], and for
asymptotically hyperbolic data [2] [3]), it is only during the past few years
that much has been learned about non constant mean
curvature solutions [8] [14].

In this paper, we discuss results which show that iterative techniques
developed by Isenberg and Moncrief to study and produce non CMC solutions 
of the constraint equations on closed manifolds [14] can be adapted to
do the same for non CMC solutions which are asymptotically hyperbolic.
A careful proof of these new asymptotically hyperbolic results is
presented elsewhere [18]. Here, we discuss a bit more informally how the 
adaptation of our techniques from the closed manifold case to the 
asymptotically hyperbolic case has been carried out. We start in 
\S2 with a very brief review of the LCBY conformal formulation  
of the Einstein constraint equations, which is the starting point for all
of our analysis. In \S3, we review what we know so far about solutions 
of the constraints on closed manifolds, emphasizing how our iterative 
technique works to verify the existence of classes of non CMC solutions
on such manifolds. 
In \S4, we specify what asymptotically hyperbolic manifolds and geometries 
are, and we discuss the weighted Sobolev and weighted H\"older spaces which
we use for our studies of the constraints on these manifolds. We also 
discuss in \S4 some key PDE analytical results concerning  certain elliptic
operators on these weighted function spaces. Then in \S5, we state our
results concerning asymptotically hyperbolic non CMC solutions of the
constraints and sketch how these results are proven using the iterative 
techniques once they are adapted to asymptotically hyperbolic geometries.
We conclude in \S6 with remarks on future directions for research.

\bigskip
\beginsection \S2 {A Brief Review of the Conformal
Formulation of the Einstein }

\noindent {\bf \hskip 22 pt  Constraint Equations}
\bigskip

The vacuum Einstein constraint equations for 
initial data $(\gg, K)$ consisting of  a Riemannian metric
$\gg$ and a symmetric $\left( ^0 _2 \right)$-tensor $K$
take the form
$$\eqalignno
{& 
\nabla_a K^a{}_b - \nabla_b \left( K^c{}_c\right) =0
& (1a)\cr
& 
R - K^{ab}K_{ab} + \left(K^c{}_c\right)^2 =0
& (1b)\cr
}$$
where $R$ is the scalar curvature for $\gg$. To produce
solutions of these equations on a given three-dimensional 
manifold $\SS^3$, as well as to study and parametrize these 
solutions, it is
very useful to reformulate equations (1) using the LCBY
conformal method, developed by Lichnerowicz, Choquet-Bruhat,
and York [9]. The idea is to split the data $(\gg, K)$ into
two parts: The first part --- the conformal data $(\ll, \ss,\tt)$
--- consists of a Riemannian metric $\ll$, a symmetric tensor
$\ss$ which is trace-free ($\ll_{ab}\ss^{ab}=0$) and divergence-free
($\nabla_{a}\ss^{ab}=0$) with respect to $\ll$, and a scalar 
function $\tt$ on $\SS^3$. The second part --- the determined data
$(W, \phi)$ --- consists of a vector field $W$ and a positive 
definite scalar function $\phi$ on $\SS^3$. Then to obtain a 
solution of the constraint equations (1), one first chooses
$(\ll, \ss, \tt)$ and one then attempts to solve the equations
$$\eqalignno
{&
\nabla _a (\kl W)^a{}_b = {2\over 3} \phi^6 \nabla_b\tau
& (2a)\cr
&
\nabla^2\phi={1\over 8} R\ \phi-{1\over 8} \left(\ss^{ab}+\kl W^{ab}\right)
\left(\ss_{ab}+\kl W_{ab}\right)\phi^{-7}+{1\over{12}}\tt^2\phi^5
& (2b)\cr
}$$
for $W$ and $\phi$. Here $\nabla$ is  the covariant derivative compatible
with $\ll$, $R$ is its scalar curvature, and $\kl $ is the conformal Killing
operator
$$\kl W_{ab} := \nabla_aW_b+\nabla_bW_a-{2\over3}\ll_{ab}\nabla_cW^c\eqno{\rm
(3)}$$
If, for some chosen set of the conformal data $(\ll,\ss,\tt)$, one can find a 
solution $(W,\phi)$ for equations (2), then the reconstituted initial data
$$\eqalignno
{&\gg_{ab}=\phi^4\ll_{ab}&
(4a)\cr
&K^{cd}=\phi^{-10}\left(\ss^{cd}+\kl 
W^{cd}\right)+{1\over3}\phi^{-4}\ll^{cd}\tt
&(4b)\cr}
$$
form a solution of the vacuum constraint equations (1).

One readily verifies that the second order operator
$\vl :W\mapsto \nabla_a(\kl W)^a{}_b$ is elliptic. Hence, in applying the
conformal reformulation to the constraint equations (1), 
one transforms them into a manifestly(nonlinear) elliptic
system of four PDEs for four unknown functions.

It is {\it not} true that for every choice of the conformal data 
$(\ll, \ss,\tt)$, there is a solution $(W, \phi)$ for equations (2). 
For example, if one works on the closed manifold $\SS^3={\bf S}^3$, 
and if one chooses $\ll$ to be a 
round sphere metric with $R=8$, one chooses $\ss$ identically zero, and one
chooses $\tt=\sqrt 8$, then the system (2) reduces to
$$\eqalignno
{&\nabla_a(\kl W)^a{}_b =0& (5a)\cr
&\nabla^2\phi = \phi-{1\over 8} (\kl W)^2\phi^{-7}+\phi^5&(5b)\cr}
$$
The former equations (5a) imply that $\kl W=0$, from which it follows that
(5b) takes the form
$$\nabla^2\phi=\phi+\phi^5\eqno(6)$$
This equation admits no positive definite solution on ${\bf S}^3$.
So there is no solution to (2) for this choice of conformal data.

For which choices of $(\ll, \ss,\tt)$ does a solution to (2) exist? 
We review
some of what is known in the next section.
\bigskip

\beginsection \S3 {A Brief Review of Existence Results 
for Solutions of the Einstein 
}

\noindent \hskip 22 pt {\bf Constraint Equations}

\bigskip

As noted in the introduction, studies of the
constraint equations have traditionally  focused on two cases:
solutions on closed manifolds, and asymptotically Euclidean solutions.
In each of these cases, the existence and the parametrization of solutions
is well-understood if the mean curvature $K^c{}_c$ is taken to be
constant; for non constant mean curvature, there is less --- but 
growing --- understanding.

\def\seed{$(\ll, \ss, \tt)$}

With the constant mean curvature condition imposed, the task of determining
which conformal data \seed\  permit equations (2) to be solved, 
and therefore
which conformal data map to  solutions of the constraint equations (1), is
simplified considerably. This is because if one chooses $\tt$ to be a 
constant --- a necessary and sufficient condition for the mean curvature
$K^c{}_c$ to be constant --- then equation (2a) becomes $\nabla_a(\kl 
W)^a{}_b =0$,
which admits $(\kl W)_{ab}=0$ as a solution. Consequently, the constraint 
equations reduce to one (semi-linear, elliptic) PDE (the ``Lichnerowicz
equation") 
$$\nabla^2\phi = {1\over8}R\ 
\phi-{1\over 8} \ss^{ab}\ss_{ab}\ \phi^{-7}+{1\over{12}}\tt^2\phi^5\eqno(7)
$$
to be solved for the positive definite function $\phi$.

In both the closed manifold and the asymptotically Euclidean cases,
necessary and sufficient conditions on the conformal data \seed\
are known for the Lichnerowicz equation to admit solutions. In both cases,
the behavior of the scalar curvature $R$ under conformal transformations
of $\ll$ is crucial. For asymptotically Euclidean data (defined via
weighted Sobolev spaces), Cantor [7] has shown that the Lichnerowicz
equation (with $\tt=0$, which must hold for asymptotically Euclidean data
if one assumes constant mean curvature)
admits a solution if and only if there exists a conformal transformation
of $\ll$ which produces $R=0$; further, Brill and Cantor [6] give an integral
condition for such a conformal transformation to exist. For closed 
manifolds, the combined work of Choquet-Bruhat, York, O'Murchadha [9], and
Isenberg [12] shows that the criteria for existence depends upon three 
things: (1) the Yamabe class\footnote * {\baselineskip=\normalbaselineskip
The Yamabe theorem [20] shows that every metric may be conformally transformed
so that the corresponding scalar curvature is constant. 
For a given metric $\ll$, one can transform to any constant, with a 
fixed sign characteristic of that metric. 
Hence $\ll$ is contained 
in a unique Yamabe class --- ${\cal Y}^+$, ${\cal Y}^0$, or  ${\cal Y}^-$ ---
depending upon this sign.} ---
${\cal Y}^+$, ${\cal Y}^0$, or  ${\cal Y}^-$ --- of $\ll$; (2) whether
$\tt$ is zero or not; and (3) whether $\ss^2$ is identically zero or not.
For example, if 
\def\yap{{\cal Y}^+}
\def\yaz{{\cal Y}^0}
\def\yan{{\cal Y}^-}
$\ll\in \yap$, $\tt\ne 0$, and $\ss^2\equiv0$, there is no solution; while
if $\ll\in \yan$, $\tt\ne 0$, and $\ss^2 \equiv \hskip -9 pt
/ \  0$, then there is a solution.
In total, there are
twelve possibilities. One finds that for those possibilities which imply
(with $\phi > 0$, and with $\ll$ conformally transformed so that $R$ is
constant) a definite sign for the right hand side of the Lichnerowicz
equation, the equation admits no solution; otherwise a solution does exist.
A careful statement and proof of these results for constant mean curvature
data on a closed manifold is given in [13].

While the story for non constant mean curvature is much less complete,
considerable progress has been made in recent years, especially for 
data on closed manifolds.
The analysis is more difficult, because in the non CMC case one must work
with the full, coupled, system (2) rather than just the Lichnerowicz 
equation (7). However, using a sequence scheme which we will describe below,
Moncrief and Isenberg have been able to show that equation (2) admits 
solutions (on a closed manifold) for each of the following classes of
conformal data: (I) $\ll\in \yan$, $\tt^2 > 0$, and 
$|\nabla \tt|< C(\ll,\ss)$, where $C(\ll,\ss)$ is a constant depending on
$\ll$ and $\ss$ [14];
(II) $\ll \in \yap$ and $|\nabla\tt|< C(\ll,\ss)$ [15]; and (III)
$\ll\in \yaz$, $\tt^2 > 0$, and $|\nabla \tt|<C(\ll,\ss)$ [15].

The sequence method is based on the semi-decoupled sequence of 
PDEs
$$ 
\nabla_a\left(\kl W_n\right)^a{}_b = {2\over3}\  
\left(\phi_{n\hskip -2 pt -\hskip -3 pt 1}\right)^6\ \nabla_{_b}  
{\tt}\eqno(8a)$$
and
$$
\nabla^2\phi_n={1\over 8} R\ \phi_n-{1\over 8} \bigl(\ss^{ab}+\left(
\kl W_n\right)^{ab}\bigr)
\bigl(\ss_{ab}+\left(\kl W_n\right)_{ab}\bigr)\left(\phi
_n\right)^{-7}+{1\over{12}}\tt^2\left(\phi_n\right)^5
\eqno(8b)$$
These equations
are semi-decoupled in the sense that if one knows 
$\phi_{n\hskip -2 pt -\hskip -3 pt 1}$, then equation (8b) is a 
(linear, elliptic) PDE for $W_n$ alone; and then once one knows
$W_n$, equation (8b) is a (quasilinear, Lichnerowicz-type) PDE
for $\phi_n$ alone. The idea is to (a) show that there is a sequence
of solutions $\bigl\{ \left(\phi_n,W_n\right)\bigr\}$ to the sequence
of equations (8); (b) show that
the sequence $\bigl\{ \left(\phi_n,W_n\right)\bigr\}$
converges to some $\bigl\{ \left(\phi_\infty,W_\infty\right)\bigr\}$; and
finally (c) show that $\bigl\{ \left(\phi_\infty, W_\infty\right)\bigr\}$
is a solution to equation (2) for the chosen set of conformal data.

To set the stage for discussing how the sequence method has been 
adapted to the study of equations (2) with asymptotically hyperbolic data,
we will describe in a bit more detail how these three steps are carried 
out for conformal data on a closed manifold. First we consider how one shows 
that the sequence $\bigl\{ \left(\phi_n,W_n\right)\bigr\}$ exists. 
The choice of $\phi_{_0}$ is free (within certain bounds; see [14]).
Once $\phi_{_0}$ is chosen, the
 vector field $W_1$ is to be obtained
from equation (8a) with $n=1$. To show that indeed the linear elliptic
equation (8a) does determine $W_1$ for an arbitrary (sufficiently smooth)
choice of $\phi_{_0}$ and $\tt$, one needs to verify that the operator
$\vl$ is invertible on the space of vector fields being 
considered. Standard elliptic theory (see, e.g., Besse [5]) shows that
if one works with either the standard Sobolev spaces $H^p_k(\SS^3)$
or the standard H\"older spaces $C^{k,\aa}(\SS^3)$ of vector fields on a 
closed manifold, then $\vl$ is invertible; so for $\phi_{_0}$
and $\tt$ in appropriate Sobolev or H\"older spaces of functions on $\SS^3$,
$W_1$ exists. Elliptic theory on closed $\SS^3$ also shows that there
exist constants $C_1$, $C_2$, $C_3$, and $C_4$ such that
\def\lll{\left\|}
\def\rrr{\right\|}
\def\norm #1 {\lll #1 \rrr} 
$$
\left\|W_1\right\|_{H^p_{k+2}} \le C_1 \left\|
\nabla\cdot \kl  W_1 \right\|_{H^p_k} + C_2 \left\| W_1\right\|_{L^1}
\eqno(9a)
$$
and
$$
\lll{W_1}\rrr _{C^{k+2,\aa}} \le C_3 \lll{\vl W_1}\rrr _{C^{k,\aa}}
+C_4\norm{W_1} _{C^0}\eqno(9b)
$$
hold; moreover, if the metric $\ll$ has no
conformal Killing vector fields, one can replace inequalities (9) by
$$
\norm W_1 _{H^p_{k+2}} \le C_5 \norm {\vl W_1} _{H^p_k}\eqno (10a)
$$
and
$$
\norm W_1 _{C^{k+2,\aa}} \le C_6 \norm {\vl W_1} _{C^{k,\aa}}\eqno(10b)
$$
for some constants $C_5$ and $C_6$. Combining these inequalities
(10) with appropriate embedding inequalities (see, e.g., [5]), together
with equation (8a), we obtain the pointwise inequality
$$
\left|\kl W_1\right| \le C \left( \max_{\SS^3}\ \phi_{_0}\right)^6\ \left(
\max_{\SS^3} |\nabla\tt|\right),\eqno(11) 
$$
which plays an important role in step (2) of the sequence method proof.

It should be clear that the same existence and regularity results hold
for values of $n$ other than $n=1$. Thus, given a sufficiently
nice $\phi_{n\hskip -2 pt -\hskip -3 pt 1}$, one obtains $W_n$ from 
equation (8a), and $W_n$ satisfies (pointwise)
$$
\left|\kl W_n\right| \le C \left( \max_{\SS^3}\ \phi_
{n\hskip -2 pt -\hskip -3 pt 1}
\right)^6\ \left(
\max_{\SS^3} |\nabla\tt|\right),\eqno(12)
$$
for some constant $C$ independent of $n$.

Now for a given vector field $W_n$, equation (8b) is the Lichnerowicz
equation, to be solved for $\phi_n$. Hence, keeping in mind that
$\tt$ is no longer constant, one may attempt to use the techniques
which work to prove existence for constant mean curvature conformal data
[13]. For conformal data satisfying the three conditions noted earlier
in this section, the sub and super solution technique readily applies, so
long as the data are contained in appropriate Sobolev and H\"older 
spaces [14] [15]. Interestingly, we have recently been able to show that
in fact for {\it any} non CMC conformal data in appropriate function
spaces, a solution $\phi_n$ for equation (8b) exists [15]. These 
results rely upon the sub and super solution theorem for equations of
the form $\nabla^2\phi = f(\phi,x)$ on closed manifolds [14].

Once it is established that the sequence
\def\seq{$\bigl\{ \left(\phi_n,W_n\right)\bigr\}$\ }
\seq exists, one needs to show that the sequence converges. The way 
that this is done, to prove the theorems on closed manifolds,
is via a contraction mapping argument, which proceeds as follows: 
First, using equation (8b) for consecutive values of $n$, we obtain
equations of the form
$$\nabla^2\left(
\phi_{n\hskip -2 pt +\hskip -2 pt 1} - \phi_n\right)
={\cal F} \left[
\phi_{n\hskip -2 pt -\hskip -3 pt 1},
\phi_n, \phi_{n\hskip -2 pt +\hskip -2 pt 1}, x\right]\eqno(13)
$$
where $\cal F$ is a nonlocal functional of
$\phi_{n\hskip -2 pt -\hskip -3 pt 1}$,
$\phi_n$,
and
$\phi_{n\hskip -2 pt +\hskip -2 pt 1}$
(see equations (45)-(46) in [14]). Next, one establishes $n$-independent
upper and lower bounds on all $\phi_n$
[in practice, this is done by finding $n$-independent
upper and lower bounds on the sequence of sub and super solutions
$\Bigl\{\Bigl((\phi_-)_n, (\phi_+)_n\Bigr)\Bigr\}$ which one uses
to prove the existence of the sequence $\phi_n$ of solutions to
equation (8b)]. Then, using these upper and lower bounds together
with the pointwise inequalities (12), one shows that (13)
can be written as
$$\nabla^2\left(
\phi_{n\hskip -2 pt +\hskip -2 pt 1} - \phi_n\right)
-{\cal G} \left[\phi_{n\hskip -2 pt +\hskip -2 pt 1} - \phi_n
\right] = {\cal H} \left[\phi_n - \phi_{n\hskip -2 pt -\hskip -3 pt 1}\right]
\eqno (14)
$$ where
$${\cal G} \left[\phi_{n\hskip -2 pt +\hskip -2 pt 1} - \phi_n
\right] \ge \Lambda \left( \phi_{n\hskip -2 pt +\hskip -2 pt 1} - \phi_n
\right) \eqno (15a)$$
for a constant $\Lambda$ which depends upon $\tt$ alone, and
$$
{\cal H} \left[\phi_n - \phi_{n\hskip -2 pt -\hskip -3 pt 1}\right]
\le \Theta \ \max_{\SS^3} \left( 
\phi_n - \phi_{n\hskip -2 pt -\hskip -3 pt 1} \right) \eqno (15b)
$$
where $\Theta$ is a constant depending on the conformal data \seed. 
Both $\Lambda$ and $\Theta$ are independent of $n$. It then follows 
from the maximum principle applied to (14) that
$$ \left| \phi_{n\hskip -2 pt +\hskip -2 pt 1} - \phi_n \right|
\le {\Theta \over \Lambda}\  \max_{\SS^3} \left|
\phi_n - \phi_{n\hskip -2 pt -\hskip -3 pt 1} \right|
\eqno(16)
$$
Hence, if the conformal data are such that ${\Theta \over \Lambda}< 1$,
then (16) defines a contraction mapping, which implies convergence of the 
sequence $\left\{ \phi_n \right\}$ to some positive function
$\phi_{\infty}$. Since $\left\{\phi_n\right\}$ converges, it follows 
immediately from the linear equation (8a) that the sequence 
$\left\{ W_n\right\}$ converges to some vector field
$\left\{ W_{\infty}\right\}$ as well.

So long as the conformal data
are chosen with a sufficiently high degree of differentiability
--- e.g., $\ll \in C^3(\SS^3),\  \ss\in H^p_2$, and $\tt\in H^p_2$
with $p>3$ --- it is fairly straightforward to show that the 
limits $\left(\phi_{\infty},W_{\infty}\right)$ of the converging sequence 
$\Bigl\{\Bigl(\phi_n, W_n\Bigr)\Bigr\}$ satisfy the constraint 
equations (2). One first shows that  $\left(\phi_{\infty},W_{\infty}\right)$
constitutes a weak solution; then one uses standard boot strap arguments 
to argue that  $\left(\phi_{\infty},W_{\infty}\right)$ are sufficiently
differentiable --- i.e., $C^2$ --- that they constitute a strong solution of
(2). This completes the proof of the existence of solutions corresponding
to certain families of conformal data on closed manifolds.
\bigskip

We would like to show that the sequence method just sketched can be 
adapted for use in producing and studying solutions of the constraints
which are asymptotically hyperbolic. Before doing this, we need to carefully
define asymptotically hyperbolic geometries and discuss the relevant 
function spaces and differential operators on these geometries. 
\bigskip

\beginsection \S4 {Analysis on Asymptotically Hyperbolic Geometries }

\bigskip

While the intuitive idea of an asymptotically hyperbolic geometry
is that of a Riemannian metric $\gg$ on a non compact manifold $\SS^3$
with $\gg$  asymptotically approaching a constant negative curvature
metric $h$ as one approaches ``infinity" on $\SS^3$, it is more useful
to work with a definition based on conformal compactification:

\bigskip

\noindent {\bf Definition 1} : {\sl
A Riemannian geometry $\left( \SS^3, \gg\right)$
is {\bf asymptotically hyperbolic} if and only if there exists a triple
$\left( \Lambda^3,\rho,\psi\right)$ where
}
{\sl\parindent 28pt
\item{a)}
$\Lambda^3$ is a smooth manifold with boundary.
\item{b)} $\rho : \Lambda^3 \rightarrow {\bf R} $ is a smooth non-negative
function, with $\rho(x) = 0$ if and only if
$x\in \partial \Lambda^3$ and  with $d\rho(x) \ne 0$ for 
$x\in \partial \Lambda^3$.
\item{c)} 
$\psi : {\rm int}\left(\Lambda^3\right) \rightarrow  \SS^3$
is a smooth diffeomorphism, with $\rho^2 \psi^{*}(\gg)$ a smooth Riemannian
metric on $ {\rm int}\left(\Lambda^3\right) $ which extends smoothly
to $\Lambda^3$. 
\item{}}

One readily verifies that if $\left( \SS^3, \gg\right)$ is asymptotically
hyperbolic in the sense of this definition, then indeed the intuitive sense
of asymptotically hyperbolic is realized. Note that the function
$\rho \circ\psi^{-1} : \SS^3 \rightarrow{\bf R}$ can be used effectively
as an ``inverse radial coordinate" which approaches zero as one moves
toward the asymptotic region on $\SS^3$. It is sometimes called the 
``defining function" for the asymptotic region.

To study differential operators like the Laplacian and $\vl$ on an 
asymptotically hyperbolic geometry 
\def\sig{$\left( \SS^3, \gg\right)$}
\sig, one needs to effectively specify boundary conditions on the 
tensor fields upon which the operators act. There are a number of ways
in which this can be done; the most useful way for our work here is
through the use of weighted H\"older and weighted Sobolev spaces. These
spaces are defined in the usual way, with the norms containing an indexed 
weight factor $\rho^{-\dd}$, where $\rho$ is the defining function
discussed above. That is, if we use $u$ to denote a covariant 
tensor field of fixed rank, $D$ to denote the $\gg$-compatible
covariant differential, and $D^j$ to denote the $j^{\rm th}$ iteration
of $D$, then the weighted H\"older spaces\footnote * 
{\baselineskip=\normalbaselineskip Note that for convenience, here we only 
define H\"older spaces with H\"older index $\aa$ being zero. More general
spaces can be defined, but are not needed here.}
 $C^\dd_k$ (for any non-negative integer $k$, and any real $\dd$) are
defined via the weighted norms
$$\|u\|_{C^\dd_k} := \sum_{j=0}^k \sup_{\SS^3}
\left|\rho^{-\dd} D^ju\right|_{\gg}\eqno(17)
$$
Similarly the weighted Sobolev spaces $H^{p,\dd}_k$ are defined via the 
weighted Sobolev norms
$$\|u\|_{H^{p,\dd}_k}:=\sum_{j=0}^k \left\| \rho^{-\dd} D^ju\right\|_{L^p}
\eqno(18)$$
where $\|\cdot\|_{L^p}$ indicates the $L^p$ norm on \sig.

The index ``$\dd$" in both $C^\dd_k$ and 
\def\sob{H^{p,\dd}_k}
\def\hol{C^\dd_k}
$\sob$ indicates the required asymptotic fall off rate for $D^j u$.
Specifically, since $\rho$ goes to zero asymptotically, the function 
$\rho^{-\dd}$ blows up asymptotically for positive $\dd$; hence
$\sup_{\SS^3}\left|\rho^{-\dd}D^ju\right|_{\gg}$ is finite for 
$\dd>0$ only if $\left| D^ju\right|_{\gg}$ goes to zero quickly
enough. Negative $\dd$ allows $\left| D^ju\right|_{\gg}$ to go to
zero more slowly, if at all. In general, larger $\dd$ means that 
a faster fall-off rate is required.

For carrying out the iteration method proof (especially the
bootstrap steps at the end) it is important to know the embedding
theorems for these function spaces, which describe how they all relate
to each other. In summary, for tensor fields on a three-dimensional
manifold, one finds [16]

\bigbreak

{\parindent 15pt
\item{a)}$\sob \subset H^{q,\ee}_l \ \ \ \  {\rm if}\
\   1\le q\le p\le\infty, 
\ \ \  \   
l\le k,  \ \ 
{\rm and}\ \ 
\dd-\ee >3\left({1\over q}-{1\over p}\right)$\hfill (19a)
\item{b)} $\hol \subset C^\ee_l $ \ \ \ \ \ \ \  \ 
if\ \  $l\le k$ \ \  and\ \   $\ee\le \dd $\hfill (19b)
\item{}\hskip -13pt   and
\item{c)}$
H^{p,\dd}_{k+s} \subset \hol$   \ \ \ \ \ \ \  \ 
if \ \  $sp>3$\hfill (19c)
\item{}}

One also has the very useful multiplication law for tensor fields
$u$ and $v$ contained in weighted subspaces:
$$\|uv\|_{H^{p,\dd+\ee}_k} \le C \|u\|_{H^{p,\dd}_k}\|v\|_{H^{p,\ee}_k}
\ \ \ \  {\rm if} \ \ kp>3\eqno(20)$$
for some constant $C$, from which it follows that if $u\in\sob$ and
$v\in H^{p,\ee}_k$ and if $kp>3$, then $uv\in H^{p,\dd+\ee}_k$.

There are four PDE analytical results which play an important role in 
the sequence method proof on closed manifolds, and which we therefore
must consider on asymptotically hyperbolic geometries if we were to extend 
this method to these geometries: the invertibility of the operators
$\vl$ and $\nabla^2$, the regularity estimates (10) for $\vl$ (leading
to the pointwise estimate (11) for $|{\rm L} W$), the maximum principle
for the Laplacian, and the sub and super solution theorem for PDEs
of the form $\nabla^2 \phi = F(\phi,x)$ --- e.g., the Lichnerowicz
equation. We will now consider each of these issues in turn.

The first two --- the invertibility and regularity  estimates for $\vl$
and $\nabla^2$ on asymptotic geometries ---
are closely tied because the proof of the first property depends upon
the validity of a version of the second; and because
once one verifies the first property, the
second one follows. Interestingly, it is only 
during this past year --- through the work of Jack Lee [17] combined with the 
earlier work of Andersson and Chru\'sciel [2] --- that these basic
results have been established to the degree of generality which we need.

\bigbreak

The explicit statement of the invertibility result is as follows:
\bigskip
\noindent
{\bf Proposition 1 }: 
{\sl
Let $1<p<\infty$ and let $k\ge 0$.

\noindent If $\left|\dd-1+{2\over p}\right|<\sqrt 3$, then
$$\vl : H^{p,\dd}_{k+2} \rightarrow \sob \eqno(21a)$$
is invertible.

\noindent If $ 0 < {\dd\over 2} + {1\over p} <1$, then
$$\nabla^2 :  H^{p,\dd}_{k+2} \rightarrow \sob \eqno(21b)$$
is invertible.
}

\bigskip

What Lee shows  [17] is that Proposition 1 holds for all values of
$p\in (1,\infty)$ so long as it holds for $p=2$.
[Note that the conditions which $\dd$ is required to satisfy in
Proposition 1 follow largely from the embedding condition (19a).]
A key step in establishing the $p=2$ result is the verification,
for the appropriate values of $\dd$ in Proposition 1,
of the ``asymptotic elliptic estimate" [17]
$$
\bigl(\ll-o(\ee)\bigr) \|u\|_{H^{2,\dd}_0(\SS_\ee)} \le \|{\cal D}u\|_
{H^{2,\dd}_0(\SS_\ee)}\eqno(22)
$$
for ${\cal D}=\nabla^2$ and ${\cal D}= \vl$; here
$\SS_\ee := \bigl\{ x\in \SS^3 |\rho(x)<\ee\bigr\}$, where
$\rho(x)$ is the defining function for the asymptotically
hyperbolic geometry, and $o(\ee)$ represents any
continuous function which vanishes as $\ee \rightarrow 0$.
From (22), one obtains [1]
$$\|u\|_{H^{2,\dd}_{k+2}} \le C \bigl( \|{\cal D}u\|_{H^{2,\dd}_{k}}
+ \|u\|_{H^{2,\dd}_k(W)}\bigr)\eqno(23)
$$
where $W$ is some compact set in $\SS^3$. Proposition 1 then
follows from (23). For more details, see [18] and the references 
cited there.

As noted, once invertibility is established for an elliptic operator,
the regularity estimate is a consequence. So, as a corollary to
Proposition 1, we have, for $p>1$ and $\left|\dd-1+{2\over p}\right| <
\sqrt 3$,
$$ \|W\|_{H^{p,\dd}_{k+2}} \le C \|\vl W\|_{\sob}
\eqno(24a)$$
and, for $p>1$ and $\left|\dd-1+{2\over p}\right| < 1$,
$$ \|\phi\|_{H^{p,\dd}_{k+2}} \le C \left\|\nabla^2\phi\right\|_{\sob}
\eqno(24b)$$

Let us now consider the maximum principle for the Laplacian
$\nabla^2$ on an asymptotically hyperbolic geometry. The maximum
principle can take a number of different forms [12]. The version 
we need says the following.

\bigskip

\noindent {\bf Proposition 2} : {\sl Let $\xi : \SS^3 \rightarrow {\bf R}$
be a positive definite continuous function with $\xi(x) \ge m >0$. 
Let $\ll : \SS^3 \rightarrow {\bf R}$ be a continuous 
function with $|\ll(x)|\le M$. If $\psi: \SS^3 \rightarrow
{\bf R} $ is a  bounded
 $C^2$ function in the interior of $\SS^3$ and if it satisfies
the equation
$$\nabla^2 \psi - \xi\psi =\ll\eqno(25a)$$
then we have 
$$|\psi|\le {M\over m}\eqno(25b)$$}

\bigskip
\noindent Note that this result follows fairly directly from a recently
proven asymptotic behavior lemma of Graham and Lee [10], together with 
the maximum principle on compact manifolds.

The remaining result we need is a sub and super solution theorem for the 
Laplacian on an asymptotically hyperbolic geometry. The result is as
follows:

\bigskip

\noindent  {\bf Proposition 3} :
{\sl  Let $0<\ee< 2\left(1-{1\over p}\right)$. Let $f$ be a functional
such that for every function $u : \SS^3 \rightarrow {\bf R}$ with 
$u-1 \in H^{p.\ee}_0\left(\SS^3\right)$, we have $f(u;\cdot)\in 
H^{p.\ee}_0\left(\SS^3\right)$. Assume that there exist a pair of functions
$\psi_- : \SS^3 \rightarrow  {\bf R}^+ $ and 
$\psi_+ : \SS^3 \rightarrow  {\bf R}^+ $ such that }
{\sl \parindent 28pt
\item{(i)} \hskip -5 pt 
\def\sub{\psi_-}
\def\supe{\psi_+}
$\sub$ and $\supe$ are both piecewise $C^2$ (i.e. they are $C^2$ outside of a
union of submanifolds of lower dimension)
\item{(ii)}
$\left(\sub -1\right) \in H^{p,\ee}_1$ and 
$\left(\supe -1\right) \in H^{p,\ee}_1$
\item{(iii)} $\sub(x) \le \supe(x)$ for all $x\in \SS^3$
\item{}\hskip -30pt and
\item{(iv)} $\nabla^2 \sub \ge f\left(\sub, x\right)$
, \ \ \ $\nabla^2 \supe \le f\left(\supe, x\right).$
\item{}}
\def\sub{\psi_-}
\def\supe{\psi_+}
{\sl 
\hskip -40 pt Then, 
there exists a  unique function $\psi : \SS^3 \rightarrow {\bf R}^+$
such that}
{
\def\sub{\psi_-}
\def\supe{\psi_+}
\sl\parindent 28pt
\item{a)} $\psi-1 \in H^{p,\ee}_3$
\item{b)} $\sub(x) \le \psi(x)\le\supe(x)$
for all $x\in \SS^3$
\item{}\hskip -30pt and
\item{c)} $\nabla^2\psi = f(\psi,x)$.}

\bigskip

The argument for proving Proposition 3 is much like
that used in the proof of Proposition 4.1 in [18]. There is one key 
extra step one needs for the result here:
At a certain point in the argument --- where one wants to show
that $\psi_1$, the first element of the sequence which will
converge to the solution $\psi$, satisfies $\psi_1 \le \supe$
--- one invokes the maximum principle. In a sense, one seems to 
need a version of the maximum priciple which would hold for 
weak solutions of (25). However, one may instead apply the $C^2$ maximum
principle (Proposition 2) in those regions where $\supe$ is $C^2$, 
and then use continuity to show that $\psi_1 \le \supe$ everywhere on $\SS^3$.
\bigskip

\beginsection \S5 {Main Result}

\bigskip

Our main result prescribes conditions on a  set of conformal data
\hfill \seed 

\noindent which are sufficient to guarantee that equations 
(2) can be solved for $\phi$ and $W$, 
and guarantee as well that the fields $(\gg, K)$
which one obtains by combining \seed and 
$(\phi, W)$ as per equations (4)
constitute (constraint-satisfying) asymptotically hyperbolic
initial data for a solution of Einstein's equations. While we have
defined above (Definition 1) what an asymptotically hyperbolic geometry
\sig \ is, we have not yet defined what asymptotically hyperbolic 
initial data $\left(\SS^3,\gg,K\right)$  are. 
The idea is that such initial data 
should correspond to the intrinsic and extrinsic 
geometry of a spacelike hypersurface which asymptotically goes to
null infinity in an asymptotically flat spacetime. One finds [2]
that the following definition is consistent with this idea:

\bigskip

\noindent {\bf Definition 2} : 
{\sl A set of initial data 
$\left(\SS^3,\gg,K\right)$ is {\bf asymptotically
hyperbolic} if 
\item{a)} $\left(\SS^3,\gg\right)$ \ is \ an \ asymptotically\  hyperbolic\ 
geometry  \ (in\  the \ sense \ of 
\item{} Definition 1)
\item{b)} ${\rm tr}_{\gg} K$  is  bounded  away  from  zero 
 asymptotically 
(i.e., outside  some $\gg$-ball,  ${\rm tr}_{\gg} K$ is non-zero)
\item{c)} The trace-free part of $K^{ab}$ is order $\rho^3$ asymptotically
\item{} (i.e., if\ \  $K^{ab}- {1\over 3} \left(
{\rm tr}_{\gg} K \right)\gg^{ab} $ is presumed
to be differentiable
to order $l$, then $K^{ab}- {1\over 3} \left(
{\rm tr}_{\gg} K\right) \gg^{ab}
\in C^{l,\dd}$ for $\dd\ge 3$).
\item{}
}

\bigskip

Now it is possible that we could choose conformal data with fairly
general asymptotic properties and then
seek solutions  $(\phi,W)$ which shift the asymptotic 
properties of the resulting initial data $(\gg,K)$ so that they match
Definition 2. However it is more straightforward to build the 
conditions of Definition 2 directly into the conformal data, and then seek
solutions $(\phi,W)$ which more or less leave these asymptotic conditions
unchanged. So we will use

\bigskip

\noindent {\bf Definition 3} :
{\sl A set of conformal data $\left( \SS^3,\ll,\ss,\tt\right)$ 
satisfy the {\bf asymptotically hyperbolic assumption} if the 
initial data $\left( \SS^3,\ll, \ss+{1\over 3}\ll\tt\right)$ are 
asymptotically hyperbolic in the sense of Definition 2.}

\bigskip

We now state our main result, which describes some additional
conditions on \seed which guarantee that we can solve (2) for $(\phi,W)$
and thereby produce an asymptotically hyperbolic solution of the 
Einstein constraint equations:

\bigskip

\noindent {\bf Theorem 1}: 
{\sl
Let $\left( \SS^3,\ll,\ss,\tt\right)$ be a set of conformal data which
satisfy the asymptotically hyperbolic assumption, plus
the following additional conditions:
\item{(i)} $\ll$ has scalar curvature $R_{\ll} < -r$ for 
some positive constant $r$.
\item{(ii)} $\ss\in H^{p,\ee}_1$ for $p>1$ and for $0<\ee<2-{2\over p}$.
\item{(iii)} $\tt$ has no zeros, \  $\tt-\sqrt{{3\over2}r} \in H^{1,\hat\ee}_p
$ for $\ee<\hat\ee<2-{2\over p}$, and $\left\|\tt - 
\sqrt{{3\over2}r} \right\|_{C^1} <\beta$ for a certain constant
$\beta$ which one can calculate from $\left(\SS^3,\ll,\ss\right)$.
\item{}
\hskip -30pt Then, there exists a unique solution $(\phi,W)$ of equations (2),
with $\phi-1 \in H^{p,\dd}_3$ for $\dd<\ee$ and $W\in H^{p,\ee}_3$. 
The resulting initial data are asymptotically hyperbolic (in the sense
of Definition 2).
}

\bigskip

This theorem has been proven using the sequence method. We now discuss
in rough terms (using the analytic results from \S4) how this works.
See [18] for a more complete discussion of the details of the proof.

\bigbreak
\def\seq{$\bigl\{ \left(\phi_n,W_n\right)\bigr\}$}

\  The first step,\  we\  recall, \ is \ to \ establish \ the\
 existence \   of\  the \ 
sequence  
 
\noindent \seq \ which satisfies the sequence of equations (8). One may choose 
$\phi_0$ freely, within bounds we will note below. It then 
follows from Proposition 1 that for  the given conformal data and for  
the values of $\ee$ hypothesized in Theorem 1, the operator $\vl$ is 
invertible,and so we obtain $W_1$.

To obtain $\phi_1$, we need to find a solution to equation (8b) with $W_1$
inserted into the right hand side. It follows from the sub and super
solution theorem (Proposition 3) that so long as we can find a 
sub solution $\left(\phi_1\right)_-$ and a 
super solution $\left(\phi_1\right)_+$ satisfying the hypotheses of 
Proposition 3, then we have $\phi_1$. 
Using the same calculations as appear in [14] for 
the closed manifold case (the hypotheses that $R_\ll$ is bounded 
negative and $\tt$ is bounded away from zero are needed here), we readily find 
{\it constants} \def\csub{\left(m_1\right)_-}\def\csup{\left(m_1\right)_+}
$\csub$ and $\csup$ which satisfy hypotheses (i) and (iii) in Proposition 3
to be sub and super solutions. However, these constants do not 
have the necessary asymptotic behavior (as required by hypotheses (ii) in
Proposition 3). To fix this, we use
$$\left(\phi_1\right)_+ := {\rm Min} \bigl\{ \csup , 1+\rho^s\bigr\}\eqno(26a)$$
and
$$\left(\phi_1\right)_- := {\rm Max} \bigl\{ \csub , 1-\rho^s\bigr\}\eqno(26b)$$
and show  (see Lemma 3.7 of [18])

\bigskip

\noindent{\bf Claim 1}: 
{\sl There exists $s>0$ such that $\left(\phi_1\right)_-$ and 
$\left(\phi_1\right)_+$ belong to $H^{p,\ee}_0$ and hence satisfy the
hypotheses of Proposition 3 to be sub and super solutions with the
desired asymptotic properties.}

\bigskip

\noindent
Note that Proposition 3 states that if one finds appropriate
sub and super solutions, then one has a unique solution to the equation of
interest. Hence we obtain $\phi_1$.

The same arguments work sequentially for all $n$, so indeed we obtain 
the sequence \seq .

We next need to verify that this sequence converges.
To do this, we rely upon a contraction mapping argument very similar to
the one used for the closed $\SS^3$ case. A key pre-requisite for the 
contraction mapping argument to work is the existence of upper and 
lower bounds on the elements of the sequence $\left\{ \phi_n\right\}$ 
which are independent of $n$. This is guaranteed by the existence
of an $n$-independent upper bound on 
$\bigl\{ \left(\phi_n\right)_+\bigr\}$ and 
 an $n$-independent lower bound on 
$\bigl\{ \left(\phi_n\right)_-\bigr\}$. Since the functions
$1+\rho^s$ and $1-\rho^s$ are bounded above and below for positive $s$ and
small $\rho$ ($\rho$ is small in the asymptotic region where $1\pm \rho^s$ 
are used), one only needs to establish $n$-independent bound on the sequences
of constants $\bigl\{ \left(m_n\right)_+\bigr\}$ and
$\bigl\{ \left(m_n\right)_+\bigr\}$. But these sequences of constants
are essentially the same as those which serve as sub and super solutions for 
$\left\{\phi_n\right\}$ in the closed manifold case. Hence the argument
used in \S5 Step 3 of [14] can be used here to establish these bounds, which 
we call $\phi_+$ and $\phi_-$. Note that the bounds within which 
$\phi_0$ must be chosen (referred to earlier) are these constants
$\phi_+$ and $\phi_-$.

Unfortunately, for a number of reasons (including the 
fact that an asymptotically hyperbolic geometry
does not have a finite volume) the rather straightforward 
calculation leading to (12) for the closed case
(see \S5 Step 1 of [14]) does not work. One can, however,
still prove the following

\bigskip

\noindent {\bf Claim 2}: 
{\sl For $p>1$, there  exists a constant $C=C(p)$
such that for all $W\in H^{p,\dd}_2$ with
$\left|\dd-1+{2\over p}\right|<\sqrt 3$, one has 
$$ |{\rm L} W | \le C\  \sup |\vl W|\eqno(27)$$}
\bigskip

The proof of this claim is fairly intricate; it proceeds as a 
proof by contradiction, with the focus being on showing that
if one could find a sequence of vector fields $V_k\in  H^{p,\dd}_2$
(with $\left|\dd-1+{2\over p}\right|<\sqrt 3$) for which
$$\sup_{\SS^3} \bigl(\left|V_k\right|+\left|{\rm L}V_k\right|\bigr)
=1\eqno(28a)$$
yet
$$\lim_{k\rightarrow\infty}\bigl[\sup\left|\vl V_k\right|\bigr] =0
\eqno(28b)$$
then one would have a contradiction. If such a sequence does not 
exist, then Claim 2 follows. It should not be a surprise that 
one could very readily produce a contradiction if we were to assume
that there exists a sequence of points $\{x_k\}$ such that
$\left|V_k(x_k)\right|+\left|{\rm L} V_k(x_k)\right|>{1\over2}$,
yet $\{x_k\}$ is contained in a compact subset of $\SS^3$. The 
much harder work comes in examining what happens if the  $\{x_k\}$
move out to infinity asymptotically. The proof works
because as one moves towards infinity, the spatial geometry
becomes (at least locally) close to a copy of a piece of hyperbolic
half-space. Details of the proof are found in Theorem 3.1 of [18].

With the $n$-independent pointwise estimate for $\left|{\rm L} W_n\right|$
established, one may proceed with the contraction mapping argument as in the
closed case. One derives equation (14)
$$\nabla^2\left(
\phi_{n\hskip -2 pt +\hskip -2 pt 1} - \phi_n\right)
-{\cal G} \left[\phi_{n\hskip -2 pt +\hskip -2 pt 1} - \phi_n
\right] = {\cal H} \left[\phi_n - \phi_{n\hskip -2 pt -\hskip -3 pt 1}\right],
\eqno (14)
$$
one establishes the estimates (15) for $\cal G$ and $\cal H$, and
then one uses the maximum principle (proposition 2) to deduce that
$$ \left| \phi_{n\hskip -2 pt +\hskip -2 pt 1} - \phi_n \right|
\le {\Theta \over \Lambda}\  \max_{\SS^3} \left|
\phi_n - \phi_{n\hskip -2 pt -\hskip -3 pt 1} \right|.
\eqno(16)
$$
Hypothesis (iii) of Theorem 1, for a certain constant $\beta$
--- see Chapter 3 of [18] --- guarantees that 
${\Theta \over \Lambda}<1$. It then follows that the sequence
$\left\{\phi_n\right\}$ converges. The convergence of the 
sequence $\left\{W_n\right\}$  immediately follows from the invertibility of 
$\vl$.

The last step of the proof of Theorem 1 involves showing that the limit
$\left(\phi_\infty,W_\infty\right)$ of the sequence \seq \  is a 
solution of the constraint equations (2), and also showing that the 
data
$
\gg_{ab}=\phi^4\ll_{ab}$
 and 
$K^{cd}=\phi^{-10}\left(\ss^{cd}+\kl
W^{cd}\right)+{1\over3}\phi^{-4}\ll^{cd}\tt$ are
asymptotically hyperbolic in the sense of Definition 2. To show that
we have a solution, we again rely on standard bootstrap
arguments. Note that while the differentiability assumptions
in Theorem 1 are weaker than we have used in the closed manifold
case (see Theorem 1 in [14]) the regularity results which we cite 
in \S4 guarantee that $\left(\phi_n -1 \right)\in C^{2,\dd}$
and $W_n\in H^{p,\ee}_2$ for $0<\dd<\ee<2-{2\over p}$. The 
contraction mapping argument for \seq only
guarantees a priori that we have $C^0$ convergence of the 
sequence. However, given this differentiability for $\left\{\phi_n\right\}$
and  $\left\{W_n\right\}$, we may bootstrap $\phi_\infty$ into 
$C^{2,\dd}$ and $W_\infty$ into  $H^{p,\ee}_4$. Thus, after using
the derivation of $\left(\phi_\infty,W_\infty\right)$ to show that
$\left(\phi_\infty,W_\infty\right)$ is a weak solution of (2), 
we can argue that they constitute a strong solution as well.

Since $\phi_\infty -1 \in C^{2,\dd}$ with $0<\dd<2-{2\over p}$, and 
$p>1$, we see that $\phi_\infty\rightarrow 1$ asymptotically. Hence, since
\sig is asymptotically hyperbolic, it follows that 
$\left(\SS^3,\phi^4\ll\right)$ is as well. 
Our assumption that $\tt$ is bounded 
away from zero by a positive constant guarantees that 
${\rm tr}_\gg K =\tt$ satisfies the second condition
in Definition 2. Finally the third condition in Definition 2 follows from our 
assumption on $\ss$ in Definition 3 together with the demonstration that
$W_\infty \in H^{p,\ee}_4$.
\bigskip

\hskip -10pt
This completes our rough sketch of the proof of our main result, Theorem 1.

\bigskip

\beginsection \S6 {Conclusion}
\bigskip

The result we discuss here --- Theorem 1 --- demonstrates the 
existence of a substantial open set of asymptotically hyperbolic
initial data which satisfy the Einstein constraint equations and
have non constant mean curvature. Theorem 1 does, however, invoke
strong restrictions on the sets of conformal data 
$\left(\SS^3, \ll,\ss,\tt\right)$ which it shows map to solutions:
1) The scalar curvature of $\ll$ must be negative.
2) The mean curvature $\tt$ must be non zero.
3) The gradient of $\tt$ is strongly controlled.

The first of these restrictions is not very severe, since it has been shown 
[4] that every asymptotically hyperbolic geometry is conformally related
to one with scalar curvature $R=-1$. One would like to remove the 
other two, however. Can one do so, and does some form of our sequence
method serve to prove the existence of solutions?

Some preliminary work indicates that we can, at least, show that the 
sequence \seq exists with these restrictions on $\tt$ removed\footnote *
{We use non constant $\left(m_n\right)_+$ and  $\left(m_n\right)_-$
in doing this.}. Whether we can then show that sequence converges is
far from clear. Work continues in this direction.

We are also interested in seeing if our method can be used to produce
non CMC asymptotically hyperbolic solutions of the constraints with
the polyhomogeneous behavior found by Andersson, Chru\'sciel and Friedrich
in the CMC case [3]. There is no reason to suspect that it cannot.

Besides these theoretical questions, we are interested in studying 
whether our method might be useful as a practical tool for producing
solutions numerically. There is interest among numerical relativists
in considering non constant mean curvature initial data. It may be that
the sequence \seq could be useful for this. 

\bigskip

\noindent {\bf Acknowledgments}

\bigskip

We thank Piotr Chru\'sciel and Jack Lee for useful
discussions pertaining to this work. We thank the 
Albert-Einstein-Institut (Potsdam) for hospitality while
portions of this work were completed. Partial support
for this research has come from NSF grant PHY 9308117 
at Oregon.
\bigskip
\medskip
\noindent{\bf References}
\medskip
\baselineskip=\normalbaselineskip
{\parindent 16 pt
\ref {[1]} L. Andersson, Elliptic system on manifolds with asymptotically
negative curvature, {\sl Indiana Univ. Math. Jour. } {\bf 42} (1993)
1359-1388

\ref {[2]} L. Andersson and P. Chru\'sciel, Solutions of the constraint 
equation in general 
relativity satisfying 
``hyperboloidal boundary conditions", {\sl Dissertationes Mathematicae} 
{\bf 355} (1996). 
.

\ref {[3]} L. Andersson, P. Chru\'sciel, and H. Friedrich, 
On the regularity of solutions to the Yamabe equation and the 
existence of smooth hyperboloidal initial data for Einstein's field
equations, {\sl Comm. Math. Phys. } {\bf 149} ( 1992), 587-612.

\ref {[4]} P. Aviles and R. McOwen, Complete conformal metrics with
negative scalar curvature in compact Riemannian manifolds,
{\sl Duke Math. J.}
{\bf 56}
(1988),
395-398

\ref {[5]} A. Besse, {\sl Einstein Manifolds}, Springer-Verlag, Berlin (1987).

\ref {[6]} D. Brill and M. Cantor, The Laplacian on asymptotically
flat manifolds and the specification of scalar curvature,
{\sl Compositio Mathematica}
{\bf 43}
(1981),
317-325.

\ref {[7]} M. Cantor, The existence of asymptotically flat initial
data for vacuum space-times, {\sl Comm. Math. Phys. } {\bf 57} (1977), 83-96.

\ref {[8]} Y. Choquet-Bruhat, J. Isenberg, and V. Moncrief, Solutions
of constraints for Einstein equations, {\sl C. R. Acad. Sci. Paris},
{\bf 315} (1992), 349-355.

\ref {[9]} Y. Choquet-Bruhat and J. York, The Cauchy problem,
{\sl General Relativity, A. Held ed.},  Plenum 1979.

\ref {[10]} C. R. Graham and J. M. Lee, Einstein metrics with prescribed
conformal infinity on the ball, {\sl Advances in Mathematics} {\bf 87} (1991)
186-225.

\ref {[11]} H. Friedrich, On the existence of n-geodesically complete or
future complete solutions of Einstein's field equations with smooth
asymptotic structure, {\sl Comm. Math. Phys.} {\bf 107} (1986), 587-609.

\ref {[12]} J. Isenberg, Parametrization of the space of solutions of
Einstein's equations, {\sl Phys. Rev. Lett.} {\bf 59} (1987), 2389-2392.

\ref {[13]} J. Isenberg, Constant mean curvature solutions of the 
Einstein constraint equations on closed manifolds, 
{\sl Class. Quantum Grav.} {\bf 12} (1995), 2249-2274.

\ref {[14]} J. Isenberg and V. Moncrief, A set of non constant mean curvature
solutions of the Einstein constraint equations on closed manifolds, 
to appear in
{\sl Class. Quantum Grav.} (1996).

\ref {[15]} J. Isenberg and V. Moncrief, { unpublished}.

\ref {[16]} A. Kufner, {\sl  Weighted Sobolev Spaces}, Wiley, New York (1985).

\ref {[17]} J. M. Lee, Fredholm operators and Einstein metrics on 
conformally compact manifolds,
in preparation. 

\ref {[18]} J. Park, Hyperboloidal non-CMC solutions  
of the Einstein constraint equations, {\sl dissertation},
University of Oregon,
(1996).

\ref {[19]} M. Protter and H. Weinberger,
{\sl Maximum Principles in Differential Equations},
Springer-Verlag, New York (1985).

\ref{ [20]} R. Schoen, Conformal deformation of a Riemannian metric to
constant scalar curvature, {\sl J. Diff. Geom.} {\bf 20} (1984), 479-495.

}
\bye